\documentclass[12pt]{article}
\usepackage{a4wide,amsmath,amssymb,graphicx}
\numberwithin{equation}{section}
\newcommand{\D}{\rlap{\hspace{0.2em}/}D}
\DeclareMathOperator{\Tr}{Tr}
\newcommand{\MS}{\ensuremath{\overline{\text{MS}}}}
\title{Decoupling in QED and QCD%
\thanks{Lecture at 5-th Helmholtz international summer school
\textit{Calculations for modern and future colliders},
Dubna, July 23 -- August 2, 2012.}}
\author{Andrey Grozin\\
Budker Institute of Nuclear Physics SB RAS, Novosibirsk\\
and Novosibirsk State University}
\date{}

\begin{document}
\maketitle
\begin{abstract}
Decoupling of a heavy flavour in QCD is discussed in a pedagogical way.
First we consider a simpler case: decoupling of muons in QED.
All calculations are done up to 2 loops.
\end{abstract}

\section{Introduction}
\label{S:Intro}

QCD with all 6 flavours is rarely used.
If characteristic momenta $p_i\ll M_Q$
(where $M_Q$ is the mass of a heavy flavour $Q$),
it is better to use a low-energy effective theory without $Q$.
Its Lagrangian has the QCD form plus $1/M_Q^n$ corrections.
Operators in the full QCD are expanded in $1/M_Q$
via appropriate operators in the effective theory.

The pioneering paper~\cite{BW:82} discussed decoupling effects
in the \MS{} scheme at two loops;
however, it contained a calculational error.
The correct two-loop result was obtained in~\cite{LRV:95}
as a by-product of a three-loop calculation.
A simple and efficient method to find decoupling effects
was developed in~\cite{CKS:98};
the two-loop result~\cite{LRV:95} was confirmed,
and new three-loop results were derived.
After that, the erratum to~\cite{BW:82} appeared,
in which the authors fixed their error,
and confirmed the results of~\cite{LRV:95,CKS:98}.
A few years ago, four-loop decoupling coefficients
have been calculated~\cite{CKS:06,SS:06}%
\footnote{The result of~\cite{SS:06} contains one master integral
which was not known analytically, only numerically, with 37-digits precision.
An analytical expression for this integral has been published later~\cite{KKOV:06}.}.

In this lecture, we shall discuss decoupling effects in QCD
at two-loop level.
First we shall discuss the QED case which is very similar to QCD but simpler
(Sect.~\ref{S:QED}).
We consider QED with electrons and muons;
when characteristic momenta $p_i \ll M$ (where $M$ is the muon mass),
the effective low-energy theory containing only electrons and photons
can be used instead.
After that we discuss QCD with light flavours $q_i$ and a heavy flavour $Q$
(Sect.~\ref{S:QCD}).

\section{Decoupling in QED}
\label{S:QED}

\subsection{Full theory and effective low-energy theory}
\label{S:QED1}

Let's consider QED with $n_f=2$ lepton flavours ---
massless electron $\psi$ and heavy muon $\Psi$:
\begin{equation}
L = \bar{\psi}_0 i \D_0 \psi_0
+ \bar{\Psi}_0 \left(i \D_0 - M_0\right) \Psi_0
- \frac{1}{4} F_{0\mu\nu} F_0^{\mu\nu}
- \frac{1}{2 a_0} \left(\partial_\mu A_0^\mu\right)^2
\label{QED:Lfull}
\end{equation}
(the covariant gauge is used; 0 means bare quantities).
When characteristic momenta $p_i \ll M$ (characteristic distances $\gg1/M$),
the low-energy effective theory containing only light fields
can be used instead.
Its Lagrangian contains all possible operators allowed by symmetries;
those with dimensionalities $>4$ are suppressed by powers of $1/M$:
\begin{equation}
L' = \bar{\psi}'_0 i \D'_0 \psi'_0
- \frac{1}{4} F'_{0\mu\nu} F_0^{\prime\mu\nu}
- \frac{1}{2 a'_0} \left(\partial_\mu A_0^{\prime\mu}\right)^2
+ \frac{1}{M_0^2} \sum_i C^0_i O^{\prime0}_i
+ \cdots
\label{QED:Leff}
\end{equation}
(quantities in the effective theory are denoted by primes).
Muons only exist in loops of size $\sim1/M$
producing local interactions of light fields.
Coefficients in the effective Lagrangian are tuned to reproduce
scattering amplitudes of the full theory
expanded in $p_i/M$ up to some order.
We shall not discuss $1/M^n$ corrections here.

Operators of the full theory can be expressed
via all operators of the effective theory which are allowed by symmetries.
Contributions of higher-dimensionality operators are suppressed
by powers of $1/M$.
Coefficients are tuned to reproduce on-shell matrix elements
of the full-theory operators expanded in $p_i/M$ up to some order.
In particular, the light fields of the full QED can be written as
\begin{equation}
\begin{split}
&A_0 = \left[\zeta_A^0\right]^{-1/2} A'_0
+ \frac{1}{M_0^2} \sum_i C^0_{Ai} O^{\prime0}_{Ai} + \cdots\\
&\psi_0 = \left[\zeta_\psi^0\right]^{-1/2} \psi'_0
+ \frac{1}{M_0^2} \sum_i C^0_{\psi i} O^{\prime0}_{\psi i} + \cdots
\end{split}
\label{QED:fields}
\end{equation}
We shall not discuss $1/M^n$ corrections.

Similarly, the parameters of the full-theory Lagrangian
are related to those of the effective-theory one:
\begin{equation}
e_0 = \left[\zeta_\alpha^0\right]^{-1/2} e'_0\,,\qquad
a_0 = \left[\zeta_A^0\right]^{-1} a'_0\,.
\label{QED:params}
\end{equation}
We shall soon see why decoupling of the gauge fixing parameter $a$
is determined by the same coefficient $\zeta_A$ as that of the photon field.

Now we shall recall some information about renormalization of QED
which will be used later.
For more details, see textbooks, e.\,g., \cite{G:07}.

The photon propagator has the structure
\begin{equation}
\begin{split}
\raisebox{0.25mm}{\includegraphics{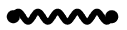}}
={}& \raisebox{0.25mm}{\includegraphics{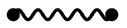}}
+ \raisebox{-3.25mm}{\includegraphics{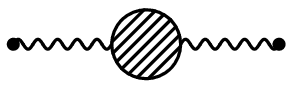}}
+ \raisebox{-3.25mm}{\includegraphics{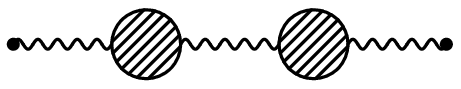}}
+ \cdots\\
-i D_{\mu\nu}(p) ={}& -i D^0_{\mu\nu}(p)
+ (-i)D^0_{\mu\alpha}(p) i\Pi^{\alpha\beta}(p) (-i)D^0_{\beta\nu}(p)\\
&{} + (-i)D^0_{\mu\alpha}(p) i\Pi^{\alpha\beta}(p) (-i) D^0_{\beta\gamma}(p)
i\Pi^{\gamma\delta}(p) (-i)D^0_{\delta\nu}(p) + \cdots
\end{split}
\label{QED:Photon}
\end{equation}
where the photon self energy $i\Pi_{\mu\nu}(p)$
is the sum of all one particle irreducible diagrams
(which cannot be cut into two disconnected pieces
by cutting a single photon line).
Due to the Ward identity, it is transverse:
\begin{equation}
\Pi_{\mu\nu}(p) = \left( p^2 g_{\mu\nu} - p_\mu p_\nu \right) \Pi(p^2)\,,
\label{QED:Pi}
\end{equation}
and we obtain
\begin{equation}
D_{\mu\nu}(p) = \frac{1}{p^2 \left[1 - \Pi(p^2)\right]}
\left( g_{\mu\nu} - \frac{p_\mu p_\nu}{p^2} \right)
+ a_0 \frac{p_\mu p_\nu}{(p^2)^2}\,.
\label{QED:D}
\end{equation}

The electron propagator has the structure
\begin{equation}
\begin{split}
\raisebox{0.25mm}{\includegraphics{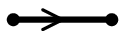}}
={}& \raisebox{0.25mm}{\includegraphics{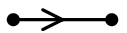}}
+ \raisebox{-3.25mm}{\includegraphics{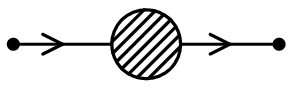}}
+ \raisebox{-3.25mm}{\includegraphics{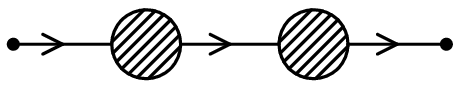}}
+ \cdots\\
i S(p) ={}& i S_0(p) + i S_0(p) (-i) \Sigma(p) i S_0(p)\\
&{} + i S_0(p) (-i) \Sigma(p) i S_0(p) (-i) \Sigma(p) i S_0(p)
+ \cdots
\end{split}
\label{QED:Electron}
\end{equation}
where the electron self energy $-i\Sigma(p)$
is the sum of all one particle irreducible diagrams
(which cannot be cut into two disconnected pieces
by cutting a single electron line).
For the massless electron, it has one Dirac structure
\begin{equation}
\Sigma(p) = \rlap/p \Sigma_V(p^2)\,,
\label{QED:Sigma}
\end{equation}
due to chirality conservation.
We obtain
\begin{equation}
S(p) = \frac{1}{\rlap/p \left[1 - \Sigma_V(p^2)\right]}\,.
\label{QED:S}
\end{equation}

We shall use two renormalization schemes, \MS{} and on-shell.
In the \MS{} scheme, bare fields and parameters are related
to renormalized ones as
\begin{equation}
\begin{split}
&A_0 = Z_A^{1/2}(\alpha(\mu))\,A(\mu)\,,\qquad
\psi_0 = Z_\psi^{1/2}(\alpha(\mu),a(\mu))\,\psi(\mu)\,,\\
&e_0 = Z_\alpha^{1/2}(\alpha(\mu))\,e(\mu)\,,\qquad
a_0 = Z_A(\alpha(\mu))\,a(\mu)\,,
\end{split}
\label{QED:MSren}
\end{equation}
where all renormalization constants have the minimal structure
\begin{equation}
Z_i(\alpha) = 1 + \frac{z_1}{\varepsilon} \frac{\alpha}{4\pi}
+ \left( \frac{z_{22}}{\varepsilon^2} + \frac{z_{21}}{\varepsilon} \right)
\left(\frac{\alpha}{4\pi}\right)^2
+ \cdots
\label{QED:min}
\end{equation}
(zeroth and positive powers of $\varepsilon$ are not allowed).
We shall see in a moment why renormalization of $a$ is determined
by the same constant $Z_A$ as that of the photon field.
In $d=4-2\varepsilon$ dimensions, dimensionality of $e$ is $\varepsilon$;
$\alpha$ should be dimensionless, therefore, we have to introduce
renormalization scale $\mu$ to construct
\begin{equation}
\frac{\alpha(\mu)}{4\pi} =
\mu^{-2\varepsilon} \frac{e^2(\mu)}{(4\pi)^{d/2}} e^{-\gamma\varepsilon}\,.
\label{QED:MSbar}
\end{equation}

Multiplying the bare photon propagator~(\ref{QED:D}) by $Z_A^{-1}$
converts $D_\bot(p^2)$ to $D^r_\bot(p^2;\mu)$ which is finite at $\varepsilon\to0$;
at the same time it converts $a_0$ to $a(\mu)$ which is also finite.
This is the reason why renormalization of the photon field $A$
and the gauge parameter $a$ is given by a single renormalization constant $Z_A$.

The sum of one-particle-irreducible vertex diagrams
(not including the external propagators) is the vertex function
\begin{equation}
\raisebox{-6mm}{\begin{picture}(30,22)
\put(15,10){\makebox(0,0){\includegraphics{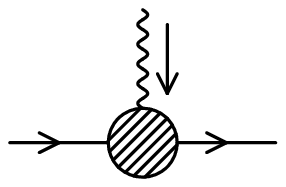}}}
\put(6.5,0){\makebox(0,0)[b]{$p$}}
\put(23.5,0){\makebox(0,0)[b]{$p'$}}
\put(20,13){\makebox(0,0)[l]{$q$}}
\put(15,22){\makebox(0,0)[t]{$\mu$}}
\end{picture}}
=i e_0 \Gamma^\mu(p,p')\,.
\label{QED:Gamma}
\end{equation}
It should be equal to $\Gamma^\mu(p,p') = Z_\Gamma \Gamma_r^\mu(p,p';\mu)$
where $Z_\Gamma$ is a minimal renormalization constant
and $\Gamma_r^\mu(p,p';\mu)$ is finite at $\varepsilon\to0$.
The scattering amplitude
$e_0 \Gamma^\mu Z_\psi Z_A^{1/2} = e \Gamma_r^\mu Z_\alpha^{1/2} Z_\Gamma Z_\psi Z_A^{1/2}$
must be finite;
this means that the product $Z_\alpha^{1/2} Z_\Gamma Z_\psi Z_A^{1/2}$ is finite.
But the only minimal~(\ref{QED:min}) renormalization constant
finite at $\varepsilon\to0$ is 1.
Therefore, $Z_\alpha^{1/2} Z_\Gamma Z_\psi Z_A^{1/2} = 1$, and
\begin{equation}
Z_\alpha = \left[Z_\Gamma Z_\psi\right]^{-2} Z_A^{-1}\,.
\label{QED:Zalpha}
\end{equation}

The Ward identity says that
\begin{equation}
\Gamma^\mu(p,p')\,q_\mu = S^{-1}(p') - S^{-1}(p)\,.
\label{QED:Ward}
\end{equation}
Multiplying this equation by $Z_\psi$ make its right-hand side finite;
hence its left-hand side is finite, too, and $Z_\Gamma = Z_\psi^{-1}$.
Thus the Ward identity makes situation in QED simple: $Z_\alpha = Z_A^{-1}$.

In the on-shell renormalization scheme
\begin{equation}
\begin{split}
&A_0 = \left[Z_A^{\text{os}}(e_0)\right]^{1/2} A_{\text{os}}\,,\qquad
\psi_0 = \left[Z_\psi^{\text{os}}(e_0,a_0)\right]^{1/2} \psi_{\text{os}}\,,\\
&e_0 = \left[Z_\alpha^{\text{os}}(e_0)\right]^{1/2} e_{\text{os}}\,,\qquad
a_0 = Z_A^{\text{os}}(e_0)\,a_{\text{os}}
\end{split}
\label{QED:os}
\end{equation}
($Z_i^{\text{os}}$ are not minimal).
By definition, the renormalized propagators in this scheme 
tend to the free ones near their mass shells:
\begin{equation}
D_\bot^{\text{os}}(p^2) \to D_\bot^0(p^2) = \frac{1}{p^2}\,,\qquad
S_{\text{os}}(p) \to S_0(p) = \frac{1}{\rlap/p}
\label{QED:Dos}
\end{equation}
at $p^2\to0$.
This means
\begin{equation}
Z_A^{\text{os}}(e_0) = \frac{1}{1 - \Pi(0)}\,,\qquad
Z_\psi^{\text{os}}(e_0,a_0) = \frac{1}{1 - \Sigma_V(0)}\,.
\label{QED:Zos}
\end{equation}

When $p$ and $p'$ are on the mass shell,
and the initial electron and the final one have physical polarizations,
a single-photon scattering of an electron is described by 2 form factors;
only one survives at $q\to0$.
The scattering amplitude in this limit is
$e_{\text{os}} \gamma^\mu = e_0 \Gamma^\mu Z_\psi^{\text{os}} \left[Z_A^{\text{os}}(e_0)\right]^{1/2}$,
where $e_{\text{os}}$ is, by definition, the on-shell electron charge.
The vertex is $\Gamma^\mu = Z_\Gamma^{\text{os}} \gamma^\mu$,
and therefore
\begin{equation}
Z_\alpha^{\text{os}} = \left[ Z_\Gamma^{\text{os}} Z_\psi^{\text{os}} \right]^{-2} \left[Z_A^{\text{os}}\right]^{-1}\,.
\label{QED:Zaos}
\end{equation}
The Ward identity~(\ref{QED:Ward}) at $q\to0$
\begin{equation}
\Gamma^\mu(p,p) = \frac{\partial S^{-1}(p)}{\partial p_\mu}
\label{QED:Ward0}
\end{equation}
near the mass shell ($p^2\to0$) gives
\begin{equation}
\Gamma^\mu(p,p) = \frac{\partial}{\partial p_\mu}
\left[\frac{Z_\psi^{\text{os}}}{\rlap/p}\right]^{-1}
= \left[Z_\psi^{\text{os}}\right]^{-1} \gamma^\mu\,,
\label{QED:Gammaos}
\end{equation}
and we obtain $Z_\Gamma^{\text{os}} = \left[Z_\psi^{\text{os}}\right]^{-1}$;
hence $Z_\alpha^{\text{os}} = \left[Z_A^{\text{os}}\right]^{-1}$.

The \MS{} renormalized fields and parameters in the full theory and the effective one are related by
\begin{equation}
\begin{split}
&A(\mu) = \zeta_{A}^{-1/2}(\mu) A'(\mu)\,,\qquad
\psi(\mu) = \zeta_{\psi}^{-1/2}(\mu) \psi'(\mu)\,,\\
&e(\mu) = \zeta_{\alpha}^{-1/2}(\mu) e'(\mu)\,,\qquad
a(\mu) = \zeta_{A}^{-1}(\mu) a'(\mu)\,,
\end{split}
\label{QED:rendec}
\end{equation}
where the renormalized decoupling coefficients are
\begin{equation}
\zeta_A(\mu) = \frac{Z_A(\alpha(\mu))}{Z'_A(\alpha'(\mu))} \zeta^0_A\,,\qquad
\zeta_\psi(\mu) = \frac{Z_\psi(\alpha(\mu),a(\mu))}{Z'_\psi(\alpha'(\mu),a'(\mu))} \zeta^0_\psi\,,\qquad
\zeta_\alpha(\mu) = \frac{Z_\alpha(\alpha(\mu))}{Z'_\alpha(\alpha'(\mu))} \zeta^0_\alpha\,.
\label{QED:zetamu}
\end{equation}
They satisfy the renormalization group equations
\begin{equation}
\begin{split}
&\frac{d\,\log\zeta_A(\mu)}{d\,\log\mu} =
\gamma_A(\alpha(\mu)) - \gamma'_A(\alpha'(\mu))\,,\\
&\frac{d\,\log\zeta_\psi(\mu)}{d\,\log\mu} =
\gamma_\psi(\alpha(\mu),a(\mu)) - \gamma'_\psi(\alpha'(\mu),a'(\mu))\,,\\
&\frac{d\,\log\zeta_\alpha(\mu)}{d\,\log\mu} =
2 \left[ \beta(\alpha(\mu)) - \beta'(\alpha'(\mu)) \right]\,,
\end{split}
\label{QED:RG}
\end{equation}
where the anomalous dimensions and the $\beta$ functions are defined as
\begin{equation}
\begin{split}
&\frac{d\,\log Z_A(\alpha(\mu))}{d\,\log\mu} = \gamma_A(\alpha(\mu))\,,\\
&\frac{d\,\log Z_\psi(\alpha(\mu),a(\mu))}{d\,\log\mu} = \gamma_\psi(\alpha(\mu),a(\mu))\,,\\
&\frac{d\,\log Z_\alpha(\alpha(\mu))}{d\,\log\mu} = 2 \beta(\alpha(\mu))\,.
\end{split}
\label{QED:gamma}
\end{equation}
Note that the \MS{} charge $\alpha(\mu)$ and the gauge parameter $a(\mu)$
satisfy the RG equations
\begin{equation}
\frac{d\,\log\alpha(\mu)}{d\,\log\mu} = - 2 \varepsilon - 2 \beta(\alpha(\mu))\,,\qquad
\frac{d\,\log a(\mu)}{d\,\log\mu} = - \gamma_A(\alpha(\mu))\,.
\label{QED:RGa}
\end{equation}

\subsection{Photon field and electron charge}
\label{S:Photon}

The propagators of both $A_{\text{os}}$ and $A'_{\text{os}}$ are equal to the free propagator at $p^2\to0$:
\begin{equation}
D_\bot^{\text{os}}(p^2) = D_\bot^{\prime\text{os}}(p^2) \left[1 + \mathcal{O}(p^2)\right]\,,
\label{Photon:D}
\end{equation}
and therefore
\begin{equation}
A_{\text{os}} = A'_{\text{os}} + \mathcal{O}\left(\frac{1}{M^2}\right)\,.
\label{Photon:A}
\end{equation}
Hence the bare decoupling coefficient is
\begin{equation}
\zeta_A^0 = \frac{Z_A^{\prime\text{os}}(e'_0)}{Z_A^{\text{os}}(e_0)}\,.
\label{Photon:zeta0}
\end{equation}
We have
\begin{equation}
Z_A^{\text{os}}(e_0) = \frac{1}{1 - \Pi(0)}\,,\qquad
Z_A^{\prime\text{os}}(e_0') = \frac{1}{1 - \Pi'(0)}\,.
\label{Photon:ZA}
\end{equation}
But $\Pi'(0) = 0$ because all loop integrals are scale-free.
Hence $Z_A^{\prime\text{os}} = 1$:
\begin{equation}
\zeta_A^0 = 1 - \Pi(0)\,.
\label{Photon:zeta0a}
\end{equation}

At one loop, the photon self energy~(\ref{QED:Pi}) is
\begin{equation*}
\raisebox{-11mm}{\begin{picture}(32,23)
\put(16,11.5){\makebox(0,0){\includegraphics{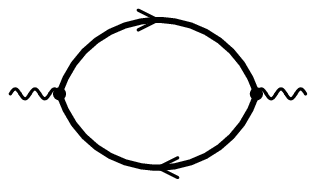}}}
\put(16,0){\makebox(0,0)[b]{$k$}}
\put(16,23){\makebox(0,0)[t]{$k+p$}}
\put(0,10.5){\makebox(0,0)[tl]{$\mu$}}
\put(32,10.5){\makebox(0,0)[tr]{$\nu$}}
\end{picture}} =
i \left( p^2 g_{\mu\nu} - p_\mu p_\nu \right) \Pi(p^2)\,.
\end{equation*}
Contraction in $\mu$ and $\nu$ we get
\begin{equation*}
\begin{split}
&\Pi(p^2) = \frac{4 i e_0^2}{(d-1) p^2}
\int \frac{d^d k}{(2\pi)^d} \frac{N}{D_1 D_2}\,,\\
&D_1 = M_0^2 - (k+p)^2\,,\qquad
D_2 = M_0^2 - k^2\,,\\
&N = \frac{1}{4} \Tr \gamma_\mu (\rlap/k+\rlap/p+M_0) \gamma^\mu (\rlap/k+M_0)\,.
\end{split}
\end{equation*}
Now we expand the integrand in $p$ up to $p^2$;
the problem reduces to trivial 1-loop vacuum integrals:
\begin{equation}
\Pi(0) = - \frac{4}{3} \frac{e_0^2 M_0^{-2\varepsilon}}{(4\pi)^{d/2}}
\Gamma(\varepsilon)\,.
\label{Photon:Pi1}
\end{equation}

The bare decoupling coefficient of the photon field with the 1-loop accuracy is
\begin{equation}
\zeta_A^0 = 1 + \frac{4}{3} \frac{e_0^2 M_0^{-2\varepsilon}}{(4\pi)^{d/2}} \Gamma(\varepsilon)
+ \cdots
\label{Photon:zeta01}
\end{equation}
Re-expressing it via renormalized quantities in the full theory, we obtain
\begin{equation}
\zeta_A^0 =
1 + \frac{4}{3} \frac{\alpha(\mu)}{4\pi\varepsilon}
Z_\alpha(\alpha(\mu)) Z_m^{-2\varepsilon}(\alpha(\mu))
\left(\frac{\mu}{M(\mu)}\right)^{2\varepsilon}
\Gamma(1+\varepsilon) e^{\gamma\varepsilon} + \cdots
\label{Photon:zeta01a}
\end{equation}
where the \MS{} renormalized muon mass is defined by
\begin{equation}
M_0 = Z_m(\alpha(\mu))\,M(\mu)\,.
\label{Photon:Zm}
\end{equation}
The \MS{} renormalization constants of the photon fields
in the full theory and the effective one with the 1-loop accuracy are
\begin{equation}
Z_A^{(\prime)}(\alpha) = 1
- \frac{4}{3} n_f \frac{\alpha}{4\pi\varepsilon} + \cdots
\label{Photon:Za1}
\end{equation}
(see, e.\,g., the textbook~\cite{G:07}).
There are 2 lepton flavours, electron and muon, in the full theory ($n_f=2$)
and only 1 (electron) in the low-energy effective theory ($n_f=1$).
We may neglect the difference between $\alpha'(\mu)$ and $\alpha(\mu)$
in corrections.
Combining these pieces together,
we arrive at the renormalized decoupling coefficient of the photon field
with the 1-loop accuracy
\begin{equation}
\zeta_A(\mu) = 1 + \frac{4}{3} L \frac{\alpha(\mu)}{4\pi} + \cdots
\label{Photon:zeta1}
\end{equation}
where
\begin{equation}
L = 2 \log \frac{\mu}{M(\mu)}\,.
\label{Photon:L}
\end{equation}
Note that $L$ depends on $\mu$ in a complicated way
because the \MS{} renormalized muon mass $M(\mu)$ also depends on $\mu$.
It is convenient to do decoupling at $\mu = \bar{M}$
where $\bar{M}$ is defined as the solution of the equation
\begin{equation}
M(\bar{M}) = \bar{M}\,;
\label{Photon:barM}
\end{equation}
then $L=0$.
The decoupling coefficient $\zeta_A(\mu)$ for other $\mu$
can be obtained by solving the RG equation~(\ref{QED:RG})
with this initial condition.

The scattering amplitude of an on-shell electron with a physical polarization
in an electromagnetic field at $q\to0$ should be the same in both theories:
\begin{equation}
e_0 \Gamma^\mu Z_\psi^{\text{os}} \left[Z_A^{\text{os}}\right]^{1/2} =
e'_0 \Gamma^{\prime\mu} Z_\psi^{\prime\text{os}} \left[Z_A^{\prime\text{os}}\right]^{1/2}\,.
\label{Photon:scattering}
\end{equation}
The vertex functions in the two theories are
\begin{equation}
\Gamma^\mu = Z_\Gamma^{\text{os}} \gamma^\mu\,,\qquad
\Gamma^{\prime\mu} = Z_\Gamma^{\prime\text{os}} \gamma^\mu\,,
\label{Photon:ZGamma}
\end{equation}
and hence
\begin{equation}
\zeta_\alpha^0
= \frac{\left[Z_\Gamma^{\text{os}} Z_\psi^{\text{os}}\right]^2 Z_A^{\text{os}}}%
{\left[Z_\Gamma^{\prime\text{os}} Z_\psi^{\prime\text{os}}\right]^2 Z_A^{\prime\text{os}}}
= \frac{Z_\alpha^{\prime\text{os}}}{Z_\alpha^{\text{os}}}\,.
\label{Photon:zetaa0a}
\end{equation}
As we already discussed, $Z_\Gamma^{\text{os}} Z_\psi^{\text{os}} = 1$ due to the Ward identity;
the same is true in the effective theory.
Moreover, in the effective theory $Z_\Gamma^{\prime\text{os}} = Z_\psi^{\prime\text{os}} = Z_A^{\prime\text{os}} = 1$
because all loop integrals are scale-free.
Therefore
\begin{equation}
\zeta_\alpha^0 = \left[\zeta_A^0\right]^{-1}\,.
\label{Photon:zetaa0}
\end{equation}

In other words, the on-shell electron charge
(measured at large distances in smooth macroscopic fields, i.\,e., at $q\to0$,
in Millikan-like experiments)
must be the same in both theories
\begin{equation}
\alpha_{\text{os}} = \alpha'_{\text{os}}\,.
\label{Photon:millikan}
\end{equation}
Thus we immediately obtain~(\ref{Photon:zetaa0a}).

\begin{figure}[ht]
\begin{center}
\includegraphics{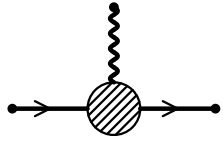}
\end{center}
\caption{The Green function of $\bar{\psi}_0$, $\psi_0$, $A_0$.}
\label{F:Green}
\end{figure}

We can look at the charge decoupling from a slightly different point of view.
Let's consider the Green function of $\bar{\psi}_0$, $\psi_0$, $A_0$.
It is the vertex with the full propagators attached (Fig.~\ref{F:Green}).
It differs from the similar Green function in the effective theory
by the bare decoupling constants of the three fields:
\begin{equation}
e_0 \Gamma S S D =
\left[\zeta_\psi^0\right]^{-1} \left[\zeta_A^0\right]^{-1/2}
e'_0 \Gamma' S' S' D'\,.
\label{Photon:Green}
\end{equation}
On the other hand,
$S = \left[\zeta_\psi^0\right]^{-1} S'$,
$D = \left[\zeta_A^0\right]^{-1} D'$, and hence
\begin{equation}
e_0 \Gamma^\mu = \zeta_\psi^0 \left[\zeta_A^0\right]^{1/2}
e'_0 \Gamma^{\prime\mu}\,.
\label{Photon:Vert}
\end{equation}
Writing down $\Gamma^\mu = \left[\zeta_\Gamma^0\right]^{-1} \Gamma^{\prime\mu}$
where $\zeta_\Gamma^0 = Z_\Gamma^{\prime\text{os}}/Z_\Gamma^{\text{os}}$,
we obtain
\begin{equation}
\zeta_\alpha^0 =
\left[\zeta_\Gamma^0 \zeta_\psi^0\right]^{-2} \left[\zeta_A^0\right]^{-1}\,,
\label{Photon:zetaa0g}
\end{equation}
and this is equivalent to~(\ref{Photon:zetaa0a}).

The \MS{} renormalized charge decoupling is
\begin{equation}
\alpha(\mu) = \zeta_\alpha^{-1}(\mu) \alpha'(\mu)\,,
\label{Photon:amu}
\end{equation}
where
\begin{equation}
\zeta_\alpha(\mu) = \frac{Z_\alpha(\alpha(\mu))}{Z'_\alpha(\alpha'(\mu))} \zeta^0_\alpha\,.
\label{Photon:zetamu}
\end{equation}
Taking into account $Z_\alpha^{(\prime)} = \bigl[Z_A^{(\prime)}\bigr]^{-1}$, we obtain
\begin{equation}
\zeta_\alpha(\mu) = \zeta_A^{-1}(\mu)\,.
\label{Photon:zetamua}
\end{equation}
With the 1-loop accuracy $\zeta_A(\mu)$ is given by~(\ref{Photon:zeta1}).

\begin{figure}[ht]
\begin{center}
\begin{picture}(106,17)
\put(16,8.5){\makebox(0,0){\includegraphics{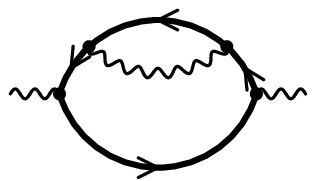}}}
\put(53,8.5){\makebox(0,0){\includegraphics{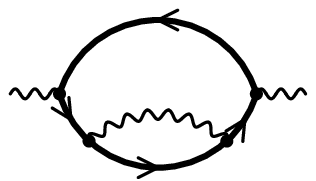}}}
\put(90,8.5){\makebox(0,0){\includegraphics{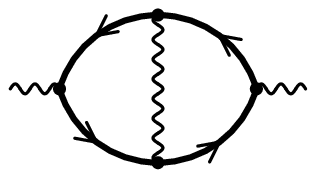}}}
\end{picture}
\end{center}
\caption{Two-loop diagrams for the photon self energy.}
\label{F:Photon}
\end{figure}

\begin{figure}[ht]
\begin{center}
\begin{picture}(22,28)
\put(11,14){\makebox(0,0){\includegraphics{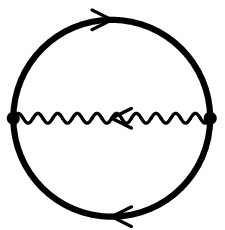}}}
\put(11,28){\makebox(0,0)[t]{$k_1$}}
\put(11,0){\makebox(0,0)[b]{$k_2$}}
\put(11,15){\makebox(0,0)[b]{$k_1-k_2$}}
\put(11,22){\makebox(0,0)[t]{$n_1$}}
\put(11,5){\makebox(0,0)[b]{$n_2$}}
\put(11,12){\makebox(0,0)[t]{$n_3$}}
\end{picture}
\end{center}
\caption{Two-loop vacuum integrals.}
\label{F:V}
\end{figure}

The photon self energy with the 2-loop accuracy (Fig.~\ref{F:Photon}) is
\begin{equation}
\begin{split}
\Pi(0) =& - \frac{4}{3} \frac{e_0^2 M_0^{-2\varepsilon}}{(4\pi)^{d/2}} \Gamma(\varepsilon)\\
&{} - \frac{2}{3} \frac{(d-4)(5d^2-33d+34)}{d(d-5)}
\left(\frac{e_0^2 M_0^{-2\varepsilon}}{(4\pi)^{d/2}}
\Gamma(\varepsilon)\right)^2 + \cdots
\end{split}
\label{Photon:Pi2}
\end{equation}
It reduces to the integrals (Fig.~\ref{F:V})
\begin{equation}
\begin{split}
&\int\frac{d^d k_1\,d^d k_2}{D_1^{n_1}D_2^{n_2}D_3^{n_3}} = -\pi^d M^{2(d-n_1-n_2-n_3)} V(n_1,n_2,n_3)\,,\\
&D_1=M^2-k_1^2\,,\qquad D_2=M^2-k_2^2\,,\qquad D_3=-(k_1-k_2)^2\,,
\end{split}
\label{Photon:Vdef}
\end{equation}
which can be expressed via $\Gamma$ functions~\cite{V:80}
\begin{equation}
V(n_1,n_2,n_3) =
\frac{\Gamma\left(\frac{d}{2}-n_3\right)
\Gamma\left(n_1+n_3-\frac{d}{2}\right)\Gamma\left(n_2+n_3-\frac{d}{2}\right)
\Gamma(n_1+n_2+n_3-d)}%
{\Gamma\left(\frac{d}{2}\right)\Gamma(n_1)\Gamma(n_2)\Gamma(n_1+n_2+2n_3-d)}\,.
\label{Photon:V}
\end{equation}

The bare decoupling coefficient of the photon field with the 2-loop accuracy is
\begin{equation}
\begin{split}
\zeta_A^0 ={}& 1
+ \frac{4}{3} \frac{e_0^2 M_0^{-2\varepsilon}}{(4\pi)^{d/2}} \Gamma(\varepsilon)\\
&{} + \frac{2}{3} \frac{(d-4)(5d^2-33d+34)}{d(d-5)}
\left(\frac{e_0^2 M_0^{-2\varepsilon}}{(4\pi)^{d/2}}
\Gamma(\varepsilon)\right)^2 + \cdots
\end{split}
\label{Photon:zeta02}
\end{equation}
Re-expressing in via renormalized quantities in the full theory, we obtain
\begin{equation}
\begin{split}
\zeta_A^0 ={}& 1
+ \frac{4}{3} e^{L\varepsilon} 
\frac{\alpha(\mu)}{4\pi\varepsilon} Z_\alpha(\alpha(\mu)) Z_m^{-2\varepsilon}(\alpha(\mu))\\
&{} - \varepsilon \left(6 - \frac{13}{3} \varepsilon + \cdots\right)
e^{2L\varepsilon} \left(\frac{\alpha(\mu)}{4\pi\varepsilon}\right)^2
+ \cdots
\end{split}
\label{Photon:zeta02a}
\end{equation}
We need 1-loop corrections to $Z_\alpha$ and $Z_m$ in the full theory:
\begin{equation}
Z_\alpha = Z_A^{-1} = 1 + 2 \cdot \frac{4}{3} \frac{\alpha(\mu)}{4\pi\varepsilon}
+ \cdots\qquad
Z_m = 1 - 3 \frac{\alpha(\mu)}{4\pi\varepsilon} + \cdots
\label{Photon:Z}
\end{equation}
The \MS{} renormalization constants of the photon fields
in the full theory and the effective one with the 2-loop accuracy are
\begin{equation}
Z_A^{(\prime)}(\alpha) = 1 - \frac{4}{3} n_f \frac{\alpha}{4\pi\varepsilon}
- 2 \varepsilon n_f \left(\frac{\alpha}{4\pi\varepsilon}\right)^2
+ \cdots
\label{Photon:Za2}
\end{equation}
(see, e.\,g., the textbook~\cite{G:07}).
Substituting everything into $\zeta_A(\mu)$~(\ref{QED:zetamu}),
we arrive at
\begin{equation}
\zeta_\alpha^{-1}(\mu) = \zeta_A(\mu) =
1 + \frac{4}{3} L \frac{\alpha(\mu)}{4\pi}
+ \left( - 4 L + \frac{13}{3} \right)
\left(\frac{\alpha(\mu)}{4\pi}\right)^2
+ \cdots
\label{Photon:zeta2}
\end{equation}
where $L$ is defined in~(\ref{Photon:L}).

In particular, at $\mu=\bar{M}$~(\ref{Photon:barM}) ($L=0$)
\begin{equation}
\zeta_\alpha(\bar{M}) = 1 - \frac{13}{3}
\left(\frac{\alpha(\bar{M})}{4\pi}\right)^2 + \cdots
\label{Photon:zetabarM}
\end{equation}
If we use $\mu = M_{\text{os}}$ instead,
where the on-shell muon mass is related to the \MS{} one by
\begin{equation}
\frac{M(\mu)}{M_{\text{os}}} = 1
- 6 \left( \log\frac{\mu}{M_{\text{os}}} + \frac{2}{3} \right)
\frac{\alpha}{4\pi} + \cdots
\label{Photon:Mos}
\end{equation}
then $L = 8 \frac{\alpha}{4\pi}$ and
\begin{equation}
\zeta_\alpha(M_{\text{os}}) = 1 - 15
\left(\frac{\alpha(M_{\text{os}})}{4\pi}\right)^2 + \cdots
\label{Photon:zetaMos}
\end{equation}
In general, for any $\mu=\bar{M}(1+\mathcal{O}(\alpha))$
we have $\zeta_\alpha(\mu) = 1 + \mathcal{O}(\alpha^2)$
(with different coefficients of $\alpha^2$);
however, for, say, $\mu = 2 \bar{M}$ or $\bar{M}/2$
the 1-loop term appears.
It is most convenient to use some $\mu_0=\bar{M}(1+\mathcal{O}(\alpha))$
as an initial condition, and obtain $\zeta_\alpha(\mu)$ for other $\mu$
from the RG equation~(\ref{QED:RG}).

\subsection{Electron field and mass}
\label{S:Electron}

The propagators of both $\psi_{\text{os}}$ and $\psi'_{\text{os}}$ are equal to the free propagator at $p^2\to0$:
\begin{equation}
\rlap/p S_{\text{os}}(p) =  \rlap/p S'_{\text{os}}(p) \left[1 + \mathcal{O}(p^2)\right]\,,
\label{Electron:SS}
\end{equation}
and therefore
\begin{equation}
\psi_{\text{os}} = \psi'_{\text{os}} + \mathcal{O}\left(\frac{1}{M^2}\right)\,.
\label{Electron:psi}
\end{equation}
Hence the bare decoupling coefficient is
\begin{equation}
\zeta_\psi^0 = \frac{Z_\psi^{\prime\text{os}}(e'_0)}{Z_\psi^{\text{os}}(e_0)}\,,
\label{Electron:zeta0}
\end{equation}
where
\begin{equation}
Z_\psi^{\text{os}}(e_0) = \frac{1}{1 - \Sigma_V(0)}\,,\qquad
Z_\psi^{\prime\text{os}}(e_0') = \frac{1}{1 - \Sigma_V'(0)}\,.
\label{Electron:Zpsi}
\end{equation}
Only diagrams with muon loops contribute to $\Sigma_V(0)$;
$\Sigma_V'(0) = 0$ because all loop integrals are scale-free.
Hence $Z_\psi^{\prime\text{os}} = 1$:
\begin{equation}
\zeta_\psi^0 = 1 - \Sigma_V(0)\,.
\label{Electron:zeta0a}
\end{equation}

\begin{figure}[ht]
\begin{center}
\includegraphics{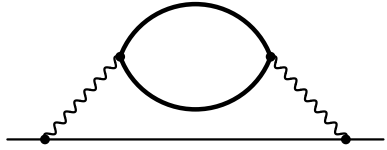}
\end{center}
\caption{Electron self energy.}
\label{F:Electron}
\end{figure}

The first diagram contributing to $\Sigma_V(0)$
appears at 2 loops (Fig.~\ref{F:Electron}):
\begin{equation*}
- i \rlap/p \Sigma_V(p^2) = \int \frac{d^d k}{(2\pi)^d}
i e_0 \gamma^\mu i \frac{\rlap/k+\rlap/p}{(k+p)^2} i e_0 \gamma^\nu
\left(\frac{-i}{k^2}\right)^2\,
i (k^2 g_{\mu\nu} - k_\mu k_\nu) \Pi(k^2)\,,
\end{equation*}
where $\Pi(k^2)$ is the muon-loop contribution to the photon self energy.
Expanding the right-hand side up to linear terms in $p$, we obtain
\begin{equation}
\Sigma_V(0) = - i e_0^2 \frac{(d-1)(d-4)}{d}
\int \frac{d^d k}{(2\pi)^d} \frac{\Pi(k^2)}{(-k^2)^2}\,.
\label{Electron:SigmaV}
\end{equation}
The calculation reduces to the integrals~(\ref{Photon:V}).
The result is
\begin{equation}
\Sigma_V(0) =
\frac{2(d-1)(d-4)(d-6)}{d(d-2)(d-5)(d-7)}
\left(\frac{e_0^2 M_0^{-2\varepsilon}}{(4\pi)^{d/2}} \Gamma(\varepsilon)\right)^2
+ \cdots
\label{Electron:SigmaV2}
\end{equation}

The bare decoupling coefficient is
\begin{equation}
\zeta_\psi^0 = 1
- \frac{2(d-1)(d-4)(d-6)}{d(d-2)(d-5)(d-7)}
\left(\frac{e_0^2 M_0^{-2\varepsilon}}{(4\pi)^{d/2}} \Gamma(\varepsilon)\right)^2
+ \cdots
= 1
- \varepsilon \left(1 - \frac{5}{6} \varepsilon + \cdots\right)
\left(\frac{\alpha}{4\pi\varepsilon}\right)^2 + \cdots
\label{Electron:zeta02}
\end{equation}
To find the renormalized decoupling coefficient $\zeta_\psi(\mu)$~(\ref{QED:zetamu})
we need the \MS{} renormalization constants $Z^{(\prime)}_\psi$.
They are determined by the anomalous dimensions
\begin{equation}
\gamma_\psi^{(\prime)}(\alpha,a) = 2 a \frac{\alpha}{4\pi}
- (4 n_f + 3) \left(\frac{\alpha}{4\pi}\right)^2 + \cdots
\label{Electron:gamma}
\end{equation}
(see, e.\,g., the textbook~\cite{G:07}).
Taking into account $\alpha'(\bar{M}) = \alpha(\bar{M}) \left[1 + \mathcal{O}(\alpha^2)\right]$,
$a'(\bar{M}) = a(\bar{M}) \left[1 + \mathcal{O}(\alpha^2)\right]$,
we see that
\begin{equation*}
\frac{Z_\psi(\alpha(\bar{M}),a(\bar{M}))}{Z_\psi'(\alpha'(\bar{M}),a'(\bar{M}))}
= 1 + \varepsilon \left(\frac{\alpha}{4\pi\varepsilon}\right)^2\,,
\end{equation*}
and hence
\begin{equation}
\zeta_\psi(\bar{M}) =
1 + \frac{5}{6} \left(\frac{\alpha(\bar{M})}{4\pi}\right)^2 + \cdots
\label{Electron:zeta2}
\end{equation}
($\zeta_\psi(\mu)$ for any $\mu\sim\bar{M}$ has the same form,
because there is no $\mathcal{O}(\alpha_s)$ term).
To find $\zeta_\psi(\mu)$ for other values of $\mu$,
the RG equation~(\ref{QED:RG}) can be used.

Until now we considered QED with massless electrons and heavy muons.
Now let's take the small electron mass $m$ into account.
The electron self energy has two Dirac structures
\begin{equation}
\Sigma(p) = \rlap/p \Sigma_V(p^2) + m_0 \Sigma_S(p^2)\,,
\label{Electron:Sigma}
\end{equation}
and the propagator is
\begin{equation}
S(p) = \frac{1}{\rlap/p - m_0 - \Sigma(p)}
= \frac{1}{1-\Sigma_V(p^2)}\;
\frac{1}{\displaystyle\rlap/p -
\frac{1+\Sigma_S(p^2)}{1-\Sigma_V(p^2)} m_0}\,.
\label{Electron:S}
\end{equation}
Near the mass shell
\begin{equation}
\frac{1}{1-\Sigma_V(p^2)} \frac{1}{\displaystyle
\rlap/p - \frac{1+\Sigma_S(p^2)}{1-\Sigma_V(p^2)} m_0}
= \frac{\left[\zeta^0_\psi\right]^{-1}}{1-\Sigma'_V(p^2)} \frac{1}{\displaystyle
\rlap/p - \frac{1+\Sigma'_S(p^2)}{1-\Sigma'_V(p^2)} m'_0}\,.
\label{Electron:Sm}
\end{equation}
We shall work in the linear approximation in $m$.
Comparing the overall factors, we reproduce~(\ref{Electron:zeta0});
comparing the denominators, we obtain
\begin{equation}
\frac{1+\Sigma_S(0)}{1-\Sigma_V(0)} m_0 =
\frac{1+\Sigma'_S(0)}{1-\Sigma'_V(0)} m'_0
\label{Electron:zetam0a}
\end{equation}
(we may set $m_0=0$ in $\Sigma_{V,S}(0)$).
The bare masses in the two theories are related by
\begin{equation}
m_0 = \left[\zeta_m^0\right]^{-1} m'_0\,;
\label{Electron:zetamdef}
\end{equation}
we obtain
\begin{equation}
\zeta_m^0 = \left[\zeta_q^0\right]^{-1} \frac{1 + \Sigma_S(0)}{1 + \Sigma'_S(0)}
= \frac{1 + \Sigma_S(0)}{1 - \Sigma_V(0)}
\label{Electron:zetam0}
\end{equation}
(because $\Sigma'_V(0) = \Sigma'_S(0) = 0$).

In other words, the on-shell electron mass is the same in both theories:
\begin{equation}
m_{\text{os}} = m'_{\text{os}}\,,
\label{Electron:mos}
\end{equation}
because it can be directly measured in experiment.
In view of $m_0 = Z_m^{\text{os}} m_{\text{os}}$ and a similar relation in the effective theory,
this leads to
\begin{equation}
\zeta_m^0 = \frac{Z_m^{\prime\text{os}}(e_0')}{Z_m^{\text{os}}(e_0)}\,.
\label{Electron:zetam0os}
\end{equation}
The on-shell electron mass renormalization constant in the full theory $Z_m^{\prime\text{os}}$
depends on two masses, $m_{\text{os}}$ and $M_{\text{os}}$;
if we neglect corrections suppressed by powers of $m_{\text{os}}^2/M_{\text{os}}^2$,
this (more physical) definition of $\zeta_m^0$
coincides with our previous definition
based on expansion in $m$ up to linear terms.

The first diagram contributing to $\Sigma_S(0)$
appears at 2 loops (Fig.~\ref{F:Electron}).
In the linear approximation, we retain $m_0$ in the numerator of the electron propagator,
and set $m_0=0$ in other places.
The result is
\begin{equation}
\Sigma_S(0) =
- \frac{2 (d-1) (d-6)}{(d-2) (d-5) (d-7)}
\left(\frac{e_0^2 M_0^{-2\varepsilon}}{(4\pi)^{d/2}} \Gamma(\varepsilon)\right)^2
+ \cdots
\label{Electron:SigmaS2}
\end{equation}

Substituting it and~(\ref{Electron:SigmaV2}) into~(\ref{Electron:zetam0}), we obtain
\begin{equation}
\begin{split}
\zeta_m^0 &{}= 1
- \frac{8 (d-1) (d-6)}{d (d-2) (d-5) (d-7)}
\left(\frac{e_0^2 M_0^{-2\varepsilon}}{(4\pi)^{d/2}} \Gamma(\varepsilon)\right)^2
+ \cdots\\
&{}= 1
+ \left(2 - \frac{5}{3} \varepsilon + \frac{89}{18} \varepsilon^2 + \cdots\right)
\left(\frac{\alpha}{4\pi\varepsilon}\right)^2 + \cdots
\end{split}
\label{Electron:zetam02}
\end{equation}
To find the renormalized decoupling coefficient
\begin{equation}
\zeta_m(\mu) = \frac{Z_m(\alpha(\mu))}{Z'_m(\alpha'(\mu))} \zeta_m^0
\label{Electron:zetammu}
\end{equation}
we need the \MS{} renormalization constants $Z^{(\prime)}_m$.
They are determined by the anomalous dimensions
\begin{equation}
\gamma_m^{(\prime)}(\alpha) = 6 \frac{\alpha}{4\pi}
+ \left(3 - \frac{20}{3} n_f\right)
\left(\frac{\alpha}{4\pi}\right)^2 + \cdots
\label{Electron:gammam}
\end{equation}
(see, e.\,g., the textbook~\cite{G:07});
we obtain
\begin{equation*}
\frac{Z_m(\alpha(\bar{M}))}{Z'_m(\alpha'(\bar{M}))} = 1
- \left(2 - \frac{5}{3} \varepsilon\right)
\left(\frac{\alpha}{4\pi\varepsilon}\right)^2 + \cdots
\end{equation*}
and hence
\begin{equation}
\zeta_m(\bar{M}) = 1 + \frac{89}{18} \left(\frac{\alpha(\bar{M})}{4\pi}\right)^2
+ \cdots
\label{Electron:zetammu2}
\end{equation}
To find $\zeta_m(\mu)$ for other values of $\mu$,
the RG equation
\begin{equation}
\frac{d\log\zeta_m(\mu)}{d\log\mu}
+ \gamma_m(\alpha(\mu)) - \gamma'_m(\alpha'(\mu)) = 0
\label{Electron:RG}
\end{equation}
can be used.

\section{Decoupling in QCD}
\label{S:QCD}

\subsection{Full theory and effective low-energy theory}
\label{S:QCD1}

Now we shall consider QCD with $n_l$ massless flavours $q_i$
and a single heavy flavour $Q$:
\begin{equation}
L = \sum_{i=1}^{n_l} \bar{q}_{0i} i \D_0 q_{0i}
+ \bar{Q}_0 \left(i \D_0 - M_0\right) Q_0
- \frac{1}{4} G^a_{0\mu\nu} G^{0a\mu\nu}
- \frac{1}{2 a_0} \left(\partial_\mu A_0^{a\mu}\right)^2
+ \left(\partial_\mu \bar{c}_0^a\right) \left(D_0^\mu c_0^a\right)\,.
\label{QCD:L}
\end{equation}
When characteristic momenta $p_i \ll M$,
the low-energy effective theory containing only light fields can be used instead:
\begin{equation}
L' = \sum_{i=1}^{n_l} \bar{q}'_{0i} i \D'_0 q'_{0i}
- \frac{1}{4} G^{\prime a}_{0\mu\nu} G^{\prime0a\mu\nu}
- \frac{1}{2 a'_0} \left(\partial_\mu A_0^{\prime a\mu}\right)^2
+ \left(\partial_\mu \bar{c}_0^{\prime a}\right) \left(D_0^{\prime\mu} c_0^{\prime a}\right)
+ \mathcal{O}\left(\frac{1}{M^2}\right)\,.
\label{QCD:Leff}
\end{equation}
The fields and the parameters of the full theory are related to those
of the effective theory:
\begin{equation}
\begin{split}
&A_0 = \left[\zeta_A^0\right]^{-1/2} A'_0\,,\qquad
q_0 = \left[\zeta_q^0\right]^{-1/2} q'_0\,,\qquad
c_0 = \left[\zeta_c^0\right]^{-1/2} c'_0\,,\\
&g_0 = \left[\zeta_\alpha^0\right]^{-1/2} g'_0\,,\qquad
a_0 = \left[\zeta_A^0\right]^{-1} a'_0\,.
\end{split}
\label{QCD:dec}
\end{equation}

\begin{figure}[ht]
\begin{center}
\begin{picture}(106,39)
\put(16,30.5){\makebox(0,0){\includegraphics{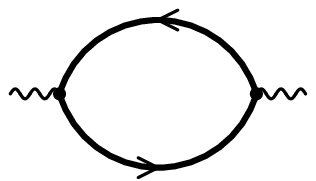}}}
\put(53,30.5){\makebox(0,0){\includegraphics{pa.eps}}}
\put(90,30.5){\makebox(0,0){\includegraphics{pc.eps}}}
\put(16,8.5){\makebox(0,0){\includegraphics{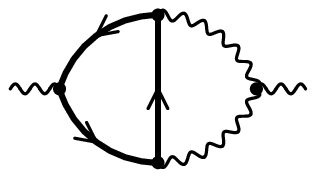}}}
\put(53,8.5){\makebox(0,0){\includegraphics{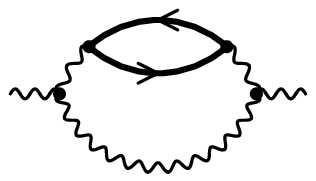}}}
\put(90,8.5){\makebox(0,0){\includegraphics{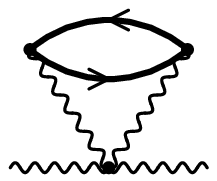}}}
\end{picture}
\end{center}
\caption{Gluon self energy up to 2 loops.}
\label{F:Gluon}
\end{figure}

The gluon self energy up to 2 loops (Fig.~\ref{F:Gluon}) is
\begin{equation}
\begin{split}
&\Pi(0) =
- \frac{4}{3} T_F \frac{g_0^2 M_0^{-2\varepsilon}}{(4\pi)^{d/2}} \Gamma(\varepsilon)\\
&{} - \frac{1}{d (d-5)}
\biggl[ \frac{2}{3} (d-4) (5 d^2 - 33 d + 34) C_F
- \frac{d^5 - 20 d^4 + 145 d^3 - 458 d^2 + 588 d - 232}{(d-2) (d-7)}
C_A \biggr]\\
&{}\times T_F
\left(\frac{g_0^2 M_0^{-2\varepsilon}}{(4\pi)^{d/2}} \Gamma(\varepsilon)\right)^2
+ \cdots
\end{split}
\label{QCD:Pi}
\end{equation}
(the $C_F$ term trivially follows from QED;
the calculation reduces to the vacuum integrals~(\ref{Photon:V})).
The bare gluon decoupling coefficient is $\zeta_A^0=1-\Pi(0)$.

The quark self energy up to 2 loops (Fig.~\ref{F:Electron})
can be trivially obtained from the QCD results~(\ref{Electron:SigmaV2}), (\ref{Electron:SigmaS2}):
\begin{equation}
\begin{split}
&\Sigma_V(0) =
\frac{2 (d-1) (d-4) (d-6)}{d (d-2) (d-5) (d-7)}
C_F T_F \left(\frac{g_0^2 M_0^{-2\varepsilon}}{(4\pi)^{d/2}} \Gamma(\varepsilon)\right)^2
+ \cdots\\
&\Sigma_S(0) =
- \frac{2 (d-1) (d-6)}{(d-2) (d-5) (d-7)}
C_F T_F \left(\frac{g_0^2 M_0^{-2\varepsilon}}{(4\pi)^{d/2}} \Gamma(\varepsilon)\right)^2
+ \cdots
\end{split}
\label{QCD:Sigma2}
\end{equation}
Therefore, the bare decoupling coefficients for the light-quark fields~(\ref{Electron:zeta0a})
and masses~(\ref{Electron:zetam0}) are
\begin{equation}
\begin{split}
&\zeta_q^0 = 1
- \frac{2 (d-1) (d-4) (d-6)}{d (d-2) (d-5) (d-7)}
C_F T_F \left(\frac{g_0^2 M_0^{-2\varepsilon}}{(4\pi)^{d/2}} \Gamma(\varepsilon)\right)^2
+ \cdots\\
&\zeta_m^0 = 1
- \frac{8 (d-1) (d-6)}{d (d-2) (d-5) (d-7)}
C_F T_F \left(\frac{g_0^2 M_0^{-2\varepsilon}}{(4\pi)^{d/2}} \Gamma(\varepsilon)\right)^2
+ \cdots
\end{split}
\label{QCD:zetaq}
\end{equation}

\begin{figure}[ht]
\begin{center}
\includegraphics{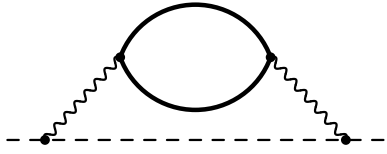}
\end{center}
\caption{Ghost self energy.}
\label{F:Ghost}
\end{figure}

The ghost propagator is
\begin{equation}
G(p) = \frac{1}{p^2 - \Sigma_c(p^2)}\,;
\label{QCD:ghost}
\end{equation}
therefore, the on-shell renormalization constant of the ghost field is
\begin{equation}
Z_c^{\text{os}} = \frac{1}{\displaystyle1 - \frac{d\Sigma_c}{d p^2}(0)}\,.
\label{QCD:Zc}
\end{equation}
The first contribution appears at 2 loops (Fig.~\ref{F:Ghost}):
\begin{equation}
\frac{d\Sigma_c}{d p^2}(0) =
- \frac{2 (d-1) (d-6)}{d (d-2) (d-5) (d-7)}
C_A T_F \left(\frac{g_0^2 M_0^{-2\varepsilon}}{(4\pi)^{d/2}} \Gamma(\varepsilon)\right)^2
+ \cdots
\label{QCD:Sigmac}
\end{equation}

\subsection{Decoupling of $\alpha_s$}
\label{S:as}

All elementary vertices of QCD are determined by a single coupling constant $g_0$.
Therefore, any vertex  (quark--gluon, 3--guon, ghost--gluon, 4--gluon)
can be used to obtain the decoupling relation for $g_0$.
In the full theory, we expand these vertex functions
in their external momenta up to the first non-vanishing term,
and compare them to the corresponding elementary vertices
in the effective theory
(there are no loop corrections in it, because there is no scale).
This first non-vanishing term of the expansion ought to have the structure
of the elementary QCD vertex,
otherwise the Lagrangian of the effective theory
would not have the QCD form
(higher terms of expansions in external momenta
lead to higher-dimensional operators in the effective low-energy Lagrangian,
suppressed by powers of $1/M^2$).

The quark--gluon vertex function at 0-th order in its external momenta
obviously has a single Dirac and colour structure $\gamma^\mu t^a$.

The 3--gluon vertex function should be expanded up to linear terms
in its external momenta.
The Bose symmetry allows two structures:
$f^{a_1 a_2 a_3} \left(g^{\mu_1 \mu_2} (k_1 - k_2)^{\mu_3} + \text{cycle}\right)$
and $d^{a_1 a_2 a_3} \left(g^{\mu_1 \mu_2} k_3^{\mu_3} + \text{cycle}\right)$.
The Slavnov--Taylor identity
${<}T\{\partial^\mu A_\mu(x),\partial^\nu A_\nu(y),\partial^\lambda A_\lambda(z)\}{>}=0$
leads to $\Gamma^{a_1 a_2 a_3}_{\mu_1 \mu_2 \mu_3} k_1^{\mu_1} k_2^{\mu_2} k_3^{\mu_3} = 0$,
thus excluding the second possibility.

The ghost--gluon vertex has a single structure: $p^\mu f^{abc}$,
where $p^\mu$ is the outgoing ghost momentum.
We shall prove this statement in a moment.

As a result, each vertex function of the full theory,
expanded in its external momenta up to the first non-vanishing term,
is equal to the corresponding elementary vertex
times a scalar quantity $\Gamma_i$.
Any one of these vertices can be used to find the bare decoupling coefficient
$\zeta_\alpha^0$:
\begin{equation}
\zeta_\alpha^0(g_0)
= \Gamma_{A\bar{c}c}^2 \left[Z_c^{\text{os}}\right]^2 Z_A^{\text{os}}
= \Gamma_{A\bar{q}q}^2 \left[Z_q^{\text{os}}\right]^2 Z_A^{\text{os}}
= \Gamma_{AAA}^2 \left[Z_A^{\text{os}}\right]^3\,.
\label{alpha:zeta0}
\end{equation}
We shall use the ghost--gluon vertex here,
because the calculation is simplest in this case.

Now we shall discuss the ghost--gluon vertex function in the full theory
expanded in its external momenta up to linear terms.
Let's consider the right-most vertex on the ghost line:
\[
\raisebox{-7mm}{\begin{picture}(27,18)
\put(13.5,9.5){\makebox(0,0){\includegraphics{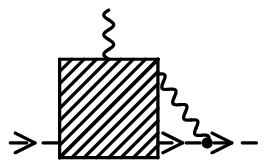}}}
\put(23.5,0){\makebox(0,0)[b]{$p$}}
\put(12.5,18){\makebox(0,0)[tl]{$\mu$}}
\put(22,4.5){\makebox(0,0)[bl]{$\nu$}}
\end{picture}}
= A^{\mu\nu} p_\nu\,.
\]
The tensor $A^{\mu\nu}$ may be calculated at zero external momenta,
hence $A^{\mu\nu}=Ag^{\mu\nu}$.
Therefore all loop diagrams have the Lorentz structure of the tree vertex,
as expected.

Now let's consider the left-most vertex:
\[
\raisebox{-7mm}{\begin{picture}(27,18)
\put(13.5,9.5){\makebox(0,0){\includegraphics{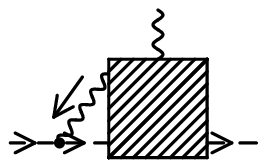}}}
\put(3.5,0){\makebox(0,0)[b]{$0$}}
\put(8.5,0){\makebox(0,0)[b]{$k$}}
\put(5,10){\makebox(0,0){$k$}}
\end{picture}}\,.
\]
It gives $k^\lambda$,
thus singling out the longitudinal part of the gluon propagator.
Therefore, all loop corrections vanish in Landau gauge.
Furthermore,
diagrams with self-energy insertions into the left-most gluon propagator
vanish in any covariant gauge: 
\[
\raisebox{-9mm}{\includegraphics{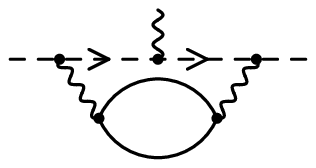}} =
\raisebox{-7mm}{\includegraphics{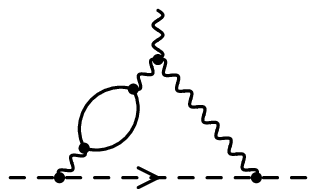}} = 0\,.
\]

In the diagrams including a quark triangle, the contraction of
$k^\lambda$ transforms
the gluon propagator to a spin 0 propagator and a factor $k^\rho$ which
contracts the quark-gluon vertex. After decomposing $\rlap/k$ into a
difference of the involved fermion denominators one obtains in
graphical form
\begin{align*}
&\raisebox{-7mm}{\includegraphics{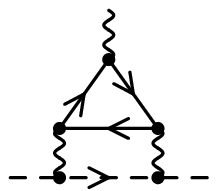}}
= a_0 \left[ \raisebox{-7mm}{\includegraphics{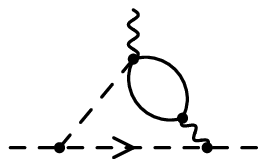}}
- \raisebox{-7mm}{\includegraphics{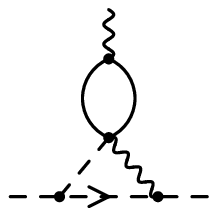}} \right]\,,\\
&\raisebox{-7mm}{\includegraphics{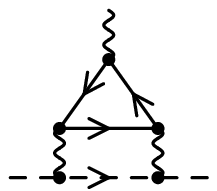}}
= a_0 \left[ \raisebox{-7mm}{\includegraphics{ghb.eps}}
- \raisebox{-7mm}{\includegraphics{gha.eps}} \right]\,.
\end{align*}
The diagrams with a massless triangle vanish.
The non-vanishing diagrams contain the same Feynman integral,
but differ by the order of the colour matrices along the quark line,
thus leading to a commutator of two colour matrices.

The remaining diagram contains a three-gluon vertex with a self energy
inserted in the right-most gluon propagator. The contraction of $k^\lambda$ 
with the three-gluon vertex cancels the gluon propagator to the right of the
three-gluon vertex:
\[
\raisebox{-7mm}{\includegraphics{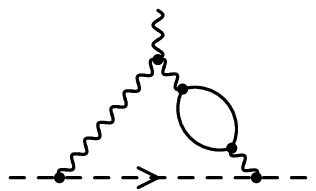}}
= a_0 \raisebox{-7mm}{\includegraphics{gha.eps}}\,.
\]
The colour structure of the three-gluon vertex is identical to the
commutator above, however with opposite sign.
Therefore, after summing all contributions the complete 2-loop result is zero.

Thus the bare decoupling coefficient $\zeta_\alpha^0$ with the 2-loop accuracy
can be obtained~(\ref{alpha:zeta0}) from the on-shell field renormalization constants
$Z_A^{\text{os}}$~(\ref{QCD:Pi}) and $Z_c^{\text{os}}$~(\ref{QCD:Sigmac}):
\begin{equation}
\begin{split}
&\left[\zeta_\alpha^0\right]^{-1} = 1
+ \frac{4}{3} T_F \frac{g_0^2 M_0^{-2\varepsilon}}{(4\pi)^{d/2}} \Gamma(\varepsilon)\\
&{} + \frac{d-4}{d (d-5)}
\left[ \frac{2}{3} (5 d^2 - 33 d + 34) C_F
- \frac{d^3 - 14 d^2 + 53 d - 32}{d-7} C_A \right]
T_F \left(\frac{g_0^2 M_0^{-2\varepsilon}}{(4\pi)^{d/2}} \Gamma(\varepsilon)\right)^2
+ \cdots
\end{split}
\label{alpha:zeta02}
\end{equation}
It is gauge invariant, as expected.
At 2 loops, $Z_A^{\text{os}}$ and $Z_c^{\text{os}}$ don't depend on $a_0$;
this means that $\Gamma_{A\bar{c}c}$ does not depend on $a_0$ at this order,
and its 2-loop term trivially vanishes in Landau gauge $a_0=0$.
Therefore non-vanishing 2-loop diagrams for $\Gamma_{A\bar{c}c}$
must cancel, as discussed above.
At 3 loops this is no longer so.

We have the expression for $g_0^{\prime2} = \zeta_\alpha^0(g_0) g_0^2$
via the bare quantities of the full theory.
We re-express it via the renormalized quantities using
\begin{equation}
\frac{g_0^2}{(4\pi)^{d/2}} \Gamma(\varepsilon) =
\mu^{2\varepsilon} \frac{\alpha_s(\mu)}{4\pi\varepsilon}
Z_\alpha(\alpha_s(\mu)) \Gamma(1+\varepsilon) e^{\gamma\varepsilon}\,,
\label{alpha:ga}
\end{equation}
where
\begin{equation}
Z_\alpha(\alpha) = 1
- \beta_0 \frac{\alpha}{4\pi\varepsilon}
+ \left(\beta_0^2 - \frac{1}{2} \beta_1 \varepsilon\right)
\left(\frac{\alpha}{4\pi\varepsilon}\right)^2 + \cdots
\label{alpha:Za}
\end{equation}
and~(\ref{Photon:Zm}).

Inverting the series
\begin{equation}
\frac{g_0^{\prime2}}{(4\pi)^{d/2}} \Gamma(\varepsilon) =
\mu^{\prime2\varepsilon} \frac{\alpha'_s(\mu')}{4\pi\varepsilon}
Z'_\alpha(\alpha'_s(\mu')) \Gamma(1+\varepsilon) e^{\gamma\varepsilon}
\label{alpha:gap}
\end{equation}
we obtain
\begin{equation}
\frac{\alpha_s'(\mu')}{4\pi\varepsilon} =
\frac{g_0^{\prime2} \mu^{\prime-2\varepsilon}}{(4\pi)^{d/2} \varepsilon} e^{-\gamma\varepsilon}
\biggl[1
+ \beta'_0 \frac{g_0^{\prime2} \mu^{\prime-2\varepsilon}}{(4\pi)^{d/2} \varepsilon} e^{-\gamma\varepsilon}
+ \left(\beta_0^{\prime2} + \frac{1}{2} \beta'_1 \varepsilon\right)
\left(\frac{g_0^{\prime2} \mu^{\prime-2\varepsilon}}{(4\pi)^{d/2} \varepsilon} e^{-\gamma\varepsilon}\right)^2
+ \cdots \biggr]\,.
\label{alpha:agp}
\end{equation}
Finally, we substitute the expression for $g_0^{\prime2}$ via $\alpha_s(\mu)$ obtained above,
and arrive at $\alpha'_s(\mu')$ expressed via via $\alpha_s(\mu)$.

\begin{figure}[ht]
\begin{center}
\begin{picture}(100,80)
\put(50,40){\makebox(0,0){\includegraphics{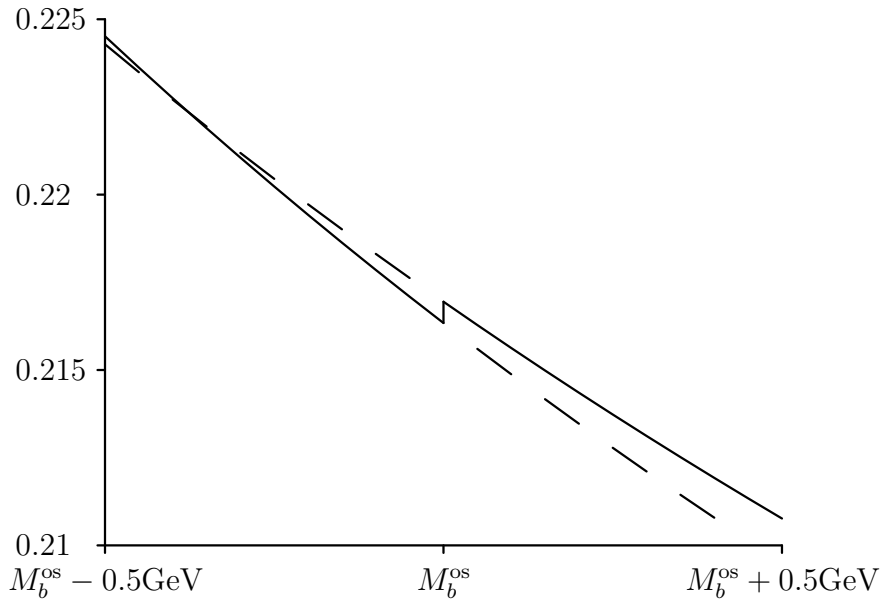}}}
\put(54,2){\makebox(0,0)[b]{$M_b^{\text{os}}$}}
\put(9,2){\makebox(0,0)[b]{$M_b^{\text{os}}-0.5\text{GeV}$}}
\put(99,2){\makebox(0,0)[b]{$M_b^{\text{os}}+0.5\text{GeV}$}}
\put(-3,9){\makebox(0,0)[l]{0.21}}
\put(-3,32.333333){\makebox(0,0)[l]{0.215}}
\put(-3,55.666667){\makebox(0,0)[l]{0.22}}
\put(-3,79){\makebox(0,0)[l]{0.225}}
\end{picture}
\end{center}
\caption{Crossing the $b$-quark threshold: $\alpha_s^{(5)}(\mu)$ and $\alpha_s^{(4)}(\mu)$.}
\label{F:alpha}
\end{figure}

The final result for $\mu'=\mu$ is $\alpha_s'(\mu) = \zeta_\alpha(\mu) \alpha_s(\mu)$, where
the renormalized decoupling coefficient is
\begin{equation}
\begin{split}
&\zeta_\alpha(\mu) = 1
- \frac{4}{3} L T_F \frac{\alpha_s(\mu)}{4\pi}\\
&{} + \biggl[ \frac{16}{9} T_F L^2
+ 4 \left(C_F - \frac{5}{3} C_A\right) L
- \left(\frac{13}{3} C_F - \frac{32}{9} C_A\right)
\biggr]
T_F \left(\frac{\alpha_s(\mu)}{4\pi}\right)^2 + \cdots
\end{split}
\label{alpha:zetamu}
\end{equation}
($L$ is defined by~(\ref{Photon:L})).
It is convenient to use $\mu=\bar{M}$ defined by~(\ref{Photon:barM}):
\begin{equation}
\zeta_\alpha(\bar{M}) = 1 
- \left( \frac{13}{3} C_F - \frac{32}{9} C_A \right) T_F
\left(\frac{\alpha_s(\bar{M})}{4\pi}\right)^2 + \cdots
\label{alpha:zetabarM}
\end{equation}
(here the $C_F$ term trivially follows from QED~(\ref{Photon:zetabarM})).
For other values of $\mu$ the RG equation~(\ref{QED:RG}) can be used.
For example, Fig.~\ref{F:alpha} (produced using \texttt{RunDec}~\cite{RunDec})
shows $\alpha_s^{(5)}(\mu)$ and $\alpha_s^{(4)}(\mu)$ near $\mu=M_b$.

The decoupling relation for the light-quark masses $m'(\bar{M}) = \zeta_m(\bar{M})\,m(\bar{M})$
can be trivially obtained from the QED result~(\ref{Electron:zetammu2}):
\begin{equation}
\zeta_m(\bar{M}) = 1
- \frac{89}{18} C_F T_F \left(\frac{\alpha_s(\bar{M})}{4\pi}\right)^2
+ \cdots
\end{equation}
For other values of $\mu$ the RG equation~(\ref{Electron:RG}) can be used.

\subsection{QCD fields}
\label{S:Gluon}

Decoupling of the gluon field and the gauge parameter
are given by the same quantity $\zeta_A^0$:
\begin{equation}
a'_0 = \zeta_A^0(g_0,a_0,M_0) a_0\,.
\label{Gluon:a0}
\end{equation}
At the first step we express the bare quantities in the right-hand side
via the renormalized ones using~(\ref{alpha:ga}), (\ref{Photon:Zm}), and
\begin{equation}
a_0 = Z_A(\alpha_s(\mu),a(\mu))\,a(\mu)\,,
\label{Gluon:a0r}
\end{equation}
and thus obtain an expression for $a'_0$
via the renormalized parameters of the full theory
$a(\mu)$, $\alpha_s(\mu)$, $M(\mu)$.
Next we find $a'(\mu')$ in terms of $a'_0$ by solving the equation
\begin{equation}
a'_0 = Z'_A(\alpha'_s(\mu'),a'(\mu'))\,a'(\mu')
\label{Gluon:renl}
\end{equation}
iteratively.
Substituting the expression for $a'_0$ obtained earlier,
we arrive at an expression for $a'(\mu')$
via $a(\mu)$, $\alpha_s(\mu)$, $M(\mu)$.

It is convenient to use $\mu'=\mu=\bar{M}$:
$a'(\bar{M}) = \zeta_A(\bar{M})\,a(\bar{M})$,
\begin{equation}
\zeta_A(\bar{M}) = 1
+ \frac{13}{12} (4 C_F - C_A) T_F \left(\frac{\alpha_s(\bar{M})}{4\pi}\right)^2
+ \cdots
\label{Gluon:zetabarM}
\end{equation}
(the $C_F$ term trivially follows from QED~(\ref{Photon:zeta2})).
Coefficients of this expansion are gauge-dependent starting from $\alpha_s^3$.
For other values of $\mu$, $\mu'$ the RG equation~(\ref{QED:RG}) can be used.

For the quark field decoupling we can trivially obtain
from the QED result~(\ref{Electron:zeta2})
\begin{equation}
\zeta_q(\bar{M}) = 1
+ \frac{5}{6} C_F T_F \left(\frac{\alpha_s(\bar{M})}{4\pi}\right)^2
+ \cdots
\label{Gluon:q}
\end{equation}
Coefficients of this expansion are gauge-dependent starting from $\alpha_s^3$.

For the ghost field we find
\begin{equation}
\zeta_c(\bar{M}) = 1
- \frac{89}{72} C_A T_F \left(\frac{\alpha_s(\bar{M})}{4\pi}\right)^2
+ \cdots
\label{Gluon:c}
\end{equation}

\section{Conclusion}
\label{S:Conc}

Decoupling effects in QED are tiny;
they are discussed in Sect.~\ref{S:QED} for pedagogical reasons:
QED calculations are similar to QCD but simpler.

In QCD decoupling effects are essential for any quantity
if we want to study it in a wide range of $\mu$.
The most important application is $\alpha_s(\mu)$:
if we want, e.\,g., to relate $\alpha_s(m_\tau)$ to $\alpha_s(m_Z)$,
we need to take into account both RG running
(in each interval it is controlled by the corresponding $\beta$ function)
and decoupling.
The same is true for, say, $m_s(\mu)$.
Parton distribution functions are also extracted from experiment
at very different values of $\mu$,
and in addition to RG running (given by the evolution equations)
we need to take into account also decoupling effects.
These effects are more difficult to calculate than, e.\,g.,
for $\alpha_s(\mu)$; however, the methods used are similar.

QCD without one or more heaviest flavours is one of the simplest examples
of an effective low-energy theory.
There are other similar examples, e.\,g.,
the effective Lagrangian of the Higgs--gluon interaction.
In the Standard Model, it is produced by the $t$-quark loop;
however, characteristic momenta are $\ll M_t$,
and this loop can be replaced by a local interaction.

I am grateful to K.\,G.~Chetyrkin, M.~H\"oschele, J.~Hoff, and M.Steinhauser
for discussions of decoupling,
and to D.\,I.~Kazakov for inviting me to give a talk at Calc-2012.
The work was supported by RFBR (grant 12-02-00106-a)
and by Russian Ministry of Education and Science.

\end{document}